\PassOptionsToPackage{unicode}{hyperref}
\PassOptionsToPackage{hyphens}{url}
\PassOptionsToPackage{dvipsnames,svgnames,x11names}{xcolor}
\documentclass[
]{article}
\usepackage{amsmath,amssymb}
\usepackage{lmodern}
\usepackage{iftex}
\ifPDFTeX
  \usepackage[T1]{fontenc}
  \usepackage[utf8]{inputenc}
  \usepackage{textcomp} 
\else 
  \usepackage{unicode-math}
  \defaultfontfeatures{Scale=MatchLowercase}
  \defaultfontfeatures[\rmfamily]{Ligatures=TeX,Scale=1}
\fi
\IfFileExists{upquote.sty}{\usepackage{upquote}}{}
\IfFileExists{microtype.sty}{
  \usepackage[]{microtype}
  \UseMicrotypeSet[protrusion]{basicmath} 
}{}
\makeatletter
\@ifundefined{KOMAClassName}{
  \IfFileExists{parskip.sty}{%
    \usepackage{parskip}
  }{
    \setlength{\parindent}{0pt}
    \setlength{\parskip}{6pt plus 2pt minus 1pt}}
}{
  \KOMAoptions{parskip=half}}
\makeatother
\usepackage{xcolor}
\usepackage{color}
\usepackage{fancyvrb}

\DefineVerbatimEnvironment{Highlighting}{Verbatim}{commandchars=\\\{\}}
\newenvironment{Shaded}{}{}

\newcommand{\CommentTok}[1]{\textcolor[rgb]{0.38,0.63,0.69}{\textit{#1}}}

\newcommand{\FloatTok}[1]{\textcolor[rgb]{0.25,0.63,0.44}{#1}}

\newcommand{\NormalTok}[1]{#1}
\newcommand{\OperatorTok}[1]{\textcolor[rgb]{0.40,0.40,0.40}{#1}}

\newcommand{\SpecialStringTok}[1]{\textcolor[rgb]{0.73,0.40,0.53}{#1}}

\newcommand{\VariableTok}[1]{\textcolor[rgb]{0.10,0.09,0.49}{#1}}

\usepackage{graphicx}
\makeatletter
\def\maxwidth{\ifdim\Gin@nat@width>\linewidth\linewidth\else\Gin@nat@width\fi}
\def\maxheight{\ifdim\Gin@nat@height>\textheight\textheight\else\Gin@nat@height\fi}
\makeatother
\setkeys{Gin}{width=\maxwidth,height=\maxheight,keepaspectratio}
\makeatletter
\def\fps@figure{htbp}
\makeatother
\providecommand{\tightlist}{%
  \setlength{\itemsep}{0pt}\setlength{\parskip}{0pt}}
\newlength{\cslhangindent}
\newlength{\csllabelwidth}
\newlength{\cslentryspacingunit} 
\newenvironment{CSLReferences}[2] 
 {
  \setlength{\parindent}{0pt}
  \ifodd #1
  \let\oldpar\par
  \def\par{\hangindent=\cslhangindent\oldpar}
  \fi
  \setlength{\parskip}{#2\cslentryspacingunit}
 }%
 {}
\usepackage{calc}

\ifLuaTeX
\usepackage[bidi=basic]{babel}
\else
\usepackage[bidi=default]{babel}
\fi
\babelprovide[main,import]{american}

\def\languageshorthands#1{}
\ifLuaTeX
  \usepackage{selnolig}  
\fi
\IfFileExists{bookmark.sty}{\usepackage{bookmark}}{\usepackage{hyperref}}
\IfFileExists{xurl.sty}{\usepackage{xurl}}{} 
\hypersetup{
  pdftitle={COBRAPRO: A MATLAB toolbox for Physics-based Battery
Modeling and Co-simulation Parameter Optimization},
  pdfauthor={Sara Ha, Simona Onori},
  pdflang={en-US},
  colorlinks=true,
  linkcolor={Maroon},
  filecolor={Maroon},
  citecolor={Blue},
  urlcolor={Blue},
  pdfcreator={LaTeX via pandoc}}

\title{COBRAPRO: A MATLAB toolbox for Physics-based Battery Modeling and
Co-simulation Parameter Optimization}

\usepackage[affil-it]{authblk}
\usepackage{orcidlink}
\author[1%
  ]{Sara Ha%
    \,\orcidlink{0009-0005-9878-3537}\,%
    }
\author[2%
  ]{Simona Onori%
    \,\orcidlink{0000-0002-6556-2608}\,%
    }

\affil[1]{Mechanical Engineering, Stanford University, 440 Escondido
Mall, Stanford 94305, CA, USA}
\affil[2]{Energy Science and Engineering, Stanford University, 367
Panama Mall, Stanford 94305, CA, USA}
\date{01 March 2024}

\begin{document}
\maketitle

\hypertarget{summary}{%
\section{Summary}\label{summary}}

COBRAPRO (\textbf{Co}-simulation \textbf{B}atte\textbf{r}y Modeling for
\textbf{A}ccelerated \textbf{P}a\textbf{r}ameter \textbf{O}ptimization)
is a physics-based battery modeling software with the capability to
perform closed-loop parameter optimization using experimental data.
COBRAPRO is based on the Doyle-Fuller-Newman (DFN) model
(\protect\hyperlink{ref-doyle_modeling_1993}{Doyle et al., 1993}), which
is the most widely-accepted high-fidelity model that considers the
lithium-ion transport and charge conservation in the liquid electrolyte
and solid electrodes, and kinetics at the solid and liquid interface
during lithium intercalation and deintercalation. Such physics-based
models have found applications in battery design
(\protect\hyperlink{ref-dai_graded_2016}{Dai \& Srinivasan, 2016}),
(\protect\hyperlink{ref-couto_lithiumion_2023}{Couto et al., 2023}) and
advanced battery management systems to ensure reliable and safe
operation of electric vehicles
(\protect\hyperlink{ref-kolluri_realtime_2020}{Kolluri et al., 2020}).
The DFN model is characterized by several physical parameters, such as
geometric, stoichiometric, concentration, transport, and kinetic
parameters, which are often unknown and need to be determined to
accurately predict battery response under various usage scenarios.
Direct measurements through cell tear-down experiments are a viable but
labor-intensive process
(\protect\hyperlink{ref-ecker_parameterization_2015a}{Ecker et al.,
2015}), (\protect\hyperlink{ref-schmalstieg_full_2018a}{Schmalstieg et
al., 2018}), (\protect\hyperlink{ref-chen_development_2020}{Chen et al.,
2020}). Furthermore, parameters obtained through experimental
characterization may necessitate further calibration to ensure
suitability for use in the DFN model
(\protect\hyperlink{ref-chen_development_2020}{Chen et al., 2020}),
since the model is a simplified representation of a real battery,
assuming perfectly spherical particles, neglecting electrode
heterogeneity, and considering internal dynamics in only one dimension.
With COBRAPRO, users can noninvasively identify parameters for any given
battery using readily available current-voltage data from a battery
cycler. COBRAPRO optimizes the DFN parameters by minimizing the error
between the simulated and experimentally observed data through an
embedded parameter optimization routine.

\hypertarget{statement-of-need}{%
\section{Statement of need}\label{statement-of-need}}

Even though parameter calibration is required to accurately predict the
dynamical behavior of actual batteries, current DFN modeling tools lack
the capability to perform parameter identification
(\protect\hyperlink{ref-sulzer_python_2021}{Sulzer et al., 2021})
(\protect\hyperlink{ref-torchio_lionsimba_2016}{Torchio et al., 2016})
(\protect\hyperlink{ref-smith_multiphase_2017}{Smith \& Bazant, 2017})
(\protect\hyperlink{ref-berliner_methods_2021}{Berliner et al., 2021}).

COMSOL Multiphysics© (\protect\hyperlink{ref-comsol}{COMSOL AB,
Stockholm, Sweden, 2023}) is a commercially available finite element
modeling software commonly used to simulate the DFN model. Although
COMSOL lacks a built-in parameter identification feature, it was
demonstrated that COMSOL's \emph{LiveLink™ for MATLAB©} can be used to
establish communication between COMSOL and MATLAB for parameter
optimization (\protect\hyperlink{ref-pozzato_general_2023}{Pozzato \&
Onori, 2023}). This framework allows users to leverage the versatile
suite of optimizers in MATLAB while running COMSOL to generate the model
output. However, the expensive licensing fee and proprietary nature of
COMSOL create barriers to public access, limiting collaboration and code
reproducibility.

In contrast, several open-source DFN model simulation tools have
emerged, such as PyBaMM
(\protect\hyperlink{ref-sulzer_python_2021}{Sulzer et al., 2021}),
LIONSIMBA (\protect\hyperlink{ref-torchio_lionsimba_2016}{Torchio et
al., 2016}), PETLION
(\protect\hyperlink{ref-berliner_methods_2021}{Berliner et al., 2021}),
DEARLIBS (\protect\hyperlink{ref-lee_robust_2021}{Lee \& Onori, 2021}),
fastDFN(\protect\hyperlink{ref-fastDFN}{Scott Moura, 2016}), and MPET
(\protect\hyperlink{ref-smith_multiphase_2017}{Smith \& Bazant, 2017}).
Among these packages, DEARLIBS is the only software equipped for
closed-loop parameter identification using experimental data. Other
packages resort to literature-derived parameter values and lack the
ability to predict real battery behavior. Taking inspiration from
DEARLIBS, COBRAPRO aims to address three primary challenges in the DFN
model:

\hypertarget{challenge-1.-computational-complexity}{%
\subsection{Challenge 1. Computational
complexity}\label{challenge-1.-computational-complexity}}

\begin{itemize}
\tightlist
\item
  \textbf{Issue:} The DFN model is also known as the
  pseudo-two-dimensional (P2D) model due to the coupling of the cell
  thickness (x-direction) and radial particle (r-direction) dimensions.
  This coupling of dimensions contributes to the model's computational
  complexity.
\item
  \textbf{Solution:} COBRAPRO leverages a fast solver to significantly
  improves the model computation speed compared to DEARLIBS. For 10
  discretized points in each domain of the cell (positive and negative
  electrodes, separator, and positive and negative active material
  particles) at 1C discharge, COBRAPRO solves the DFN model in 0.708
  seconds, while DEARLIBS takes 2.54 minutes, which is a two orders of
  magnitude improvement (\textasciitilde257 times). Under identifical
  simulation conditions, LIONSIMBA and PyBaMM computed the model in 1.13
  seconds and 0.237 seconds, respectively, demonstrating comparable
  performance to COBRAPRO. For larger discretization points, COBRAPRO
  exhibits up to three orders of magnitude improvement in computation
  speed compared to DEARLIBS.
\end{itemize}

\hypertarget{challenge-2.-consistent-initial-conditions}{%
\subsection{Challenge 2. Consistent initial
conditions}\label{challenge-2.-consistent-initial-conditions}}

\begin{itemize}
\tightlist
\item
  \textbf{Issue:} The partial differential equations (PDEs) governing
  the DFN model are discretized in x- and r-directions to form a system
  of ordinary differential equations (ODEs) and algebraic equations
  (AEs), referred to as differential-algebraic equations (DAEs). To
  solve the DAE system, the correct initial conditions of the AEs are
  required, which are typically not known \emph{a priori} for the DFN
  model. Inconsistent initial conditions result in either a failure to
  start the simulation or the model diverging towards an incorrect
  solution (\protect\hyperlink{ref-methekar_perturbation_2011}{Methekar
  et al., 2011}).
\item
  \textbf{Solution:} The single step approach
  (\protect\hyperlink{ref-lawder_extending_2015}{Lawder et al., 2015}),
  a robust initialization method, is implemented in COBRAPRO that
  automatically determines the initial conditions and seamlessly
  simulates the DFN model.
\end{itemize}

\hypertarget{challenge-3.-unknown-model-parameters}{%
\subsection{Challenge 3. Unknown model
parameters}\label{challenge-3.-unknown-model-parameters}}

\begin{itemize}
\tightlist
\item
  \textbf{Issue:} As highlighted earlier, battery parameters are
  frequently unknown, and even if obtained through experimental
  characterization, parameter calibration is essential to accurately
  simulate the DFN model.
\item
  \textbf{Solution:} A co-simulation parameter optimization framework is
  implemented in COBRAPRO. The particle swarm optimization (PSO), a
  gradient-free population-based algorithm, is employed due to its
  suitability for nonlinear models like the DFN model. The PSO aims to
  optimize parameters by minimizing the objective function defined as
  the error between the experimental and simulated voltage and state of
  charge (SOC) curves. COBRAPRO employs PSO using MATLAB's Parallel
  Computing Toolbox to accelerate PSO convergence through multicore
  processing.
\end{itemize}

\hypertarget{core-capabilities}{%
\section{Core Capabilities}\label{core-capabilities}}

\begin{itemize}
\tightlist
\item
  \textbf{Parameter identification routine:} PSO optimizes parameters
  using experimental current-voltage data
\item
  \textbf{DFN model implementation:} PDEs are discretized with finite
  volume method (FVM), and the DAE system is solved with SUNDIALS IDA
\item
  \textbf{Solid particle radial discretization options:}

  \begin{itemize}
  \tightlist
  \item
    FVM and 3rd order Hermite interpolation used to calculate particle
    surface concentration, accounting for sharp concentration gradients
    near the particle surface
    (\protect\hyperlink{ref-xu_comparative_2023}{Xu et al., 2023})
  \item
    Finite difference method (FDM)
  \end{itemize}
\item
  \textbf{DAE initialization options:}

  \begin{itemize}
  \tightlist
  \item
    Single-step approach
    (\protect\hyperlink{ref-lawder_extending_2015}{Lawder et al., 2015})
  \item
    SUNDIALS IDACalcIC
  \end{itemize}
\item
  \textbf{Simulating battery cycling:}

  \begin{itemize}
  \tightlist
  \item
    Constant current (CC) profiles
  \item
    Hybrid pulse power characterization (HPPC) profiles
  \item
    Dynamic current profiles
  \end{itemize}
\item
  \textbf{Parameter identifiability analysis:}

  \begin{itemize}
  \tightlist
  \item
    Local sensitivity analysis (LSA): Perturbs parameters around their
    nominal values and evaluates their sensitivity with respect to
    voltage and SOC
  \item
    Correlation analysis: Calculates linear correlation between two
    parameters
  \item
    Utilizes user defined sensitivity and correlation index thresholds
    to determine the set of identifiable parameters
  \end{itemize}
\end{itemize}

\hypertarget{example-case-study-on-lg-21700-m50t-cells}{%
\section{Example: Case Study on LG 21700-M50T
Cells}\label{example-case-study-on-lg-21700-m50t-cells}}

\begin{figure}
\centering
\includegraphics[width=1\textwidth,height=\textheight]{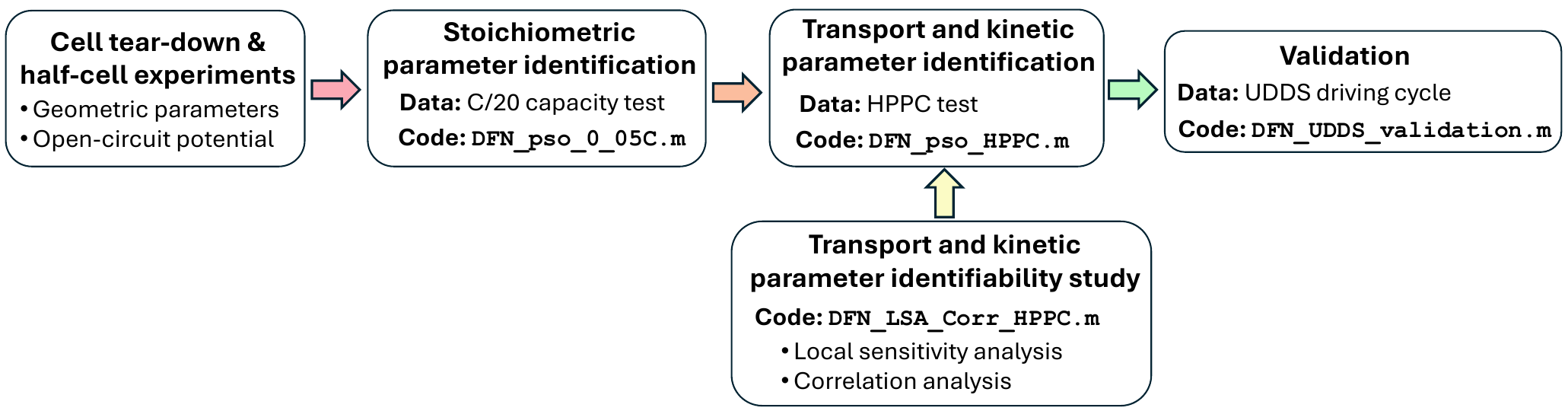}
\caption{Case study: Parameter identification procedure on LG 21700-M50T
cells.\label{fig:flowchart}}
\end{figure}

As a demonstration of COBRAPRO, we conduct a case study aimed at
parameterizing a fresh LG 21700-M50T cell using the C/20 capacity test,
HPPC, and driving cycle data
(\protect\hyperlink{ref-pozzato_data_2022}{Pozzato et al., 2022}). In
this example, we break down the identification problem by systematically
grouping parameters in each identification step, as shown in
\autoref{fig:flowchart}. This multi-step approach is proposed to improve
the identifiability of parameters instead of identifying all the unknown
parameters simultaneously
(\protect\hyperlink{ref-arunachalam_full_2019}{Arunachalam \& Onori,
2019}).

First, the geometric parameters and open-circuit potential functions are
extracted from measurements conducted in cell tear-down and half-cell
experiments on LG 21700-M50 cells, as reported by
(\protect\hyperlink{ref-chen_development_2020}{Chen et al., 2020}).
Next, the C/20 capacity test data is used to identify the stoichiometric
parameters in the example code \texttt{DFN\_pso\_0\_05C.m}. We then
conduct a parameter identifiability study, comprising of LSA and
correlation analysis, to pinpoint parameters with high sensitivity to
HPPC voltage and SOC while maintaining low correlation with other
parameters (\texttt{DFN\_LSA\_Corr\_HPPC.mat}). Next, we calibrate the
identifiable electrolyte transport and kinetic parameters using HPPC
data in the example code \texttt{DFN\_pso\_HPPC.m}. Finally, validation
of the identified parameters is carried out on the urban dynamometer
driving schedule (UDDS) data in the \texttt{DFN\_UDDS\_validation.m}
code. The \texttt{DFN\_pso\_0\_05C.m} and \texttt{DFN\_pso\_HPPC.m}
files are located in the
\texttt{Examples/Parameter\_Identification\_Routines} directory and
\texttt{DFN\_UDDS\_validation.m} is located in
\texttt{Examples/Parameter\_Identification\_Results}.

\hypertarget{c20-capacity-test-identification}{%
\subsection{C/20 Capacity Test
Identification}\label{c20-capacity-test-identification}}

In \texttt{DFN\_pso\_0\_05C.m}, the \texttt{User\ Input} section is used
to define the parameter names, the upper and lower parameter bounds for
the PSO, experimental data, etc. A preview of the \texttt{User\ Input}
section is provided here.

First, load the \texttt{Parameters\_LG\_INR21700\_M50.m} function, which
outputs a \texttt{param} structure containing the nominal DFN parameters
for a LG INR21700-M50 cell
(\protect\hyperlink{ref-chen_development_2020}{Chen et al., 2020}) and
the DFN simulation settings, e.g., discretization method, DAE
initialization method, constant or variable current type, and etc.:

\begin{Shaded}
\begin{Highlighting}[]
\CommentTok{\%\% User Input  }
\CommentTok{\% Load nominal parameters }
\VariableTok{param} \OperatorTok{=} \VariableTok{Parameters\_LG\_INR21700\_M50}\OperatorTok{;}
\end{Highlighting}
\end{Shaded}

Enter a mat file name to save PSO results. The mat file will contain an
updated \texttt{param} structure with the identified parameters from the
PSO:

\begin{Shaded}
\begin{Highlighting}[]
\CommentTok{\% Enter mat file name where your PSO results will be stored}
\VariableTok{file\_name} \OperatorTok{=} \SpecialStringTok{\textquotesingle{}identified\_parameters\_0\_05C\textquotesingle{}}\OperatorTok{;}
\end{Highlighting}
\end{Shaded}

Define the names of the parameters you want to identify. Here, we
identify the stoichiometric parameters \(\theta_p^{100}\)
(\texttt{theta100\_p}), \(\theta_n^{100}\) (\texttt{theta100\_n}),
\(\theta_p^0\) (\texttt{theta0\_p}), and \(\theta_n^0\)
(\texttt{theta0\_n}):

\begin{Shaded}
\begin{Highlighting}[]
\CommentTok{\% Enter names of parameters to identify (make sure names match the}
\CommentTok{\% parameter names in "param" structure containing the nominal parameters)}
\VariableTok{param\_CC} \OperatorTok{=}\NormalTok{ \{}\SpecialStringTok{\textquotesingle{}theta100\_p\textquotesingle{}}\OperatorTok{,} \SpecialStringTok{\textquotesingle{}theta100\_n\textquotesingle{}}\OperatorTok{,} \SpecialStringTok{\textquotesingle{}theta0\_p\textquotesingle{}}\OperatorTok{,} \SpecialStringTok{\textquotesingle{}theta0\_n\textquotesingle{}}\NormalTok{\}}\OperatorTok{;}
\end{Highlighting}
\end{Shaded}

Define the lower and upper bounds of the parameters defined in
\texttt{param\_CC}:

\begin{Shaded}
\begin{Highlighting}[]
\CommentTok{\% Enter lower and upper bounds of parameters to identify }
\CommentTok{\% theta100\_p}
\VariableTok{lower\_bounds}\NormalTok{.}\VariableTok{theta100\_p} \OperatorTok{=} \FloatTok{0.22}\OperatorTok{;} 
\VariableTok{upper\_bounds}\NormalTok{.}\VariableTok{theta100\_p} \OperatorTok{=} \FloatTok{0.34}\OperatorTok{;}
\CommentTok{\% theta100\_n}
\VariableTok{lower\_bounds}\NormalTok{.}\VariableTok{theta100\_n} \OperatorTok{=} \FloatTok{0.7}\OperatorTok{;} 
\VariableTok{upper\_bounds}\NormalTok{.}\VariableTok{theta100\_n} \OperatorTok{=} \FloatTok{1}\OperatorTok{;} 
\CommentTok{\% theta0\_p}
\VariableTok{lower\_bounds}\NormalTok{.}\VariableTok{theta0\_p} \OperatorTok{=} \FloatTok{0.7}\OperatorTok{;} 
\VariableTok{upper\_bounds}\NormalTok{.}\VariableTok{theta0\_p} \OperatorTok{=} \FloatTok{1}\OperatorTok{;} 
\CommentTok{\% theta0\_n}
\VariableTok{lower\_bounds}\NormalTok{.}\VariableTok{theta0\_n} \OperatorTok{=} \FloatTok{0.015}\OperatorTok{;} 
\VariableTok{upper\_bounds}\NormalTok{.}\VariableTok{theta0\_n} \OperatorTok{=} \FloatTok{0.04}\OperatorTok{;}
\end{Highlighting}
\end{Shaded}

Load your time, current, and voltage experimental data. In this example,
load the C/20 capacity test data:

\begin{Shaded}
\begin{Highlighting}[]
\CommentTok{\% Load Experimental Data }
\CommentTok{\%{-}{-}{-}{-}{-}{-}{-}{-}{-}{-}{-}{-}{-}{-}{-}{-}{-}{-}{-}{-}{-}{-}{-}{-}{-}{-}{-}{-}{-}{-}{-}{-}{-}{-}{-}{-}{-}{-}{-}{-}{-}{-}{-}{-}{-}{-}{-}{-}{-}{-}{-}{-}{-}{-}{-}{-}{-}{-}{-}{-}{-}{-}{-}{-}{-}{-}{-}{-}{-}{-}{-}{-}{-}{-}}
\CommentTok{\%   t: Should be a vector consisting of your time experiment data      [s] (Mx1)}
\CommentTok{\%   I: Should be a vector consisting of your current experiment data   [A] (Mx1) }
\CommentTok{\%   V: Should be a vector consisting of your voltage experiemntal data [V] (Mx1)}
\CommentTok{\%   {-}\textgreater{} where M is the total number of data points in your experiment}
\CommentTok{\%{-}{-}{-}{-}{-}{-}{-}{-}{-}{-}{-}{-}{-}{-}{-}{-}{-}{-}{-}{-}{-}{-}{-}{-}{-}{-}{-}{-}{-}{-}{-}{-}{-}{-}{-}{-}{-}{-}{-}{-}{-}{-}{-}{-}{-}{-}{-}{-}{-}{-}{-}{-}{-}{-}{-}{-}{-}{-}{-}{-}{-}{-}{-}{-}{-}{-}{-}{-}{-}{-}{-}{-}{-}{-}}
\CommentTok{\% C/20 capacity test conducted on LG INR21700 M50T cells}
\VariableTok{load}\NormalTok{(}\SpecialStringTok{\textquotesingle{}data\_INR21700\_M50T/capacity\_test\_data\_W8\_Diag1.mat\textquotesingle{}}\NormalTok{)}
\CommentTok{\% Assign your data variables to t, I, and V }
\VariableTok{t} \OperatorTok{=} \VariableTok{t\_data}\OperatorTok{;}
\VariableTok{I} \OperatorTok{=} \VariableTok{I\_data}\OperatorTok{;}
\VariableTok{V} \OperatorTok{=} \VariableTok{V\_data}\OperatorTok{;}
\end{Highlighting}
\end{Shaded}

Once the all user inputs have been defined, run the
\texttt{DFN\_pso\_0\_05C.m} code to start the PSO. Once the PSO is
finished, the code prints the identified parameter values, and the HPPC
voltage and SOC objective function values to the Command Window:

\begin{verbatim}
Displaying identified values...
------------------------
theta100_p:
Identified value: 0.26475
0.22(lower) | 0.27(initial) | 0.34(upper)
------------------------
theta100_n:
Identified value: 0.77842
0.7(lower) | 0.9014(initial) | 1(upper)
------------------------
theta0_p:
Identified value: 0.89385
0.7(lower) | 0.9084(initial) | 1(upper)
------------------------
theta0_n:
Identified value: 0.029818
0.015(lower) | 0.0279(initial) | 0.04(upper)

Displaying objective function values...
------------------------
J_V =0.0033403 [-]
J_V =11.8445 [mV]
J_SOCp =0.030231 [%]
J_SOCn =0.019037 [%]
J_tot =0.003833 [-]
\end{verbatim}

\begin{figure}
\centering
\includegraphics[width=0.65\textwidth,height=\textheight]{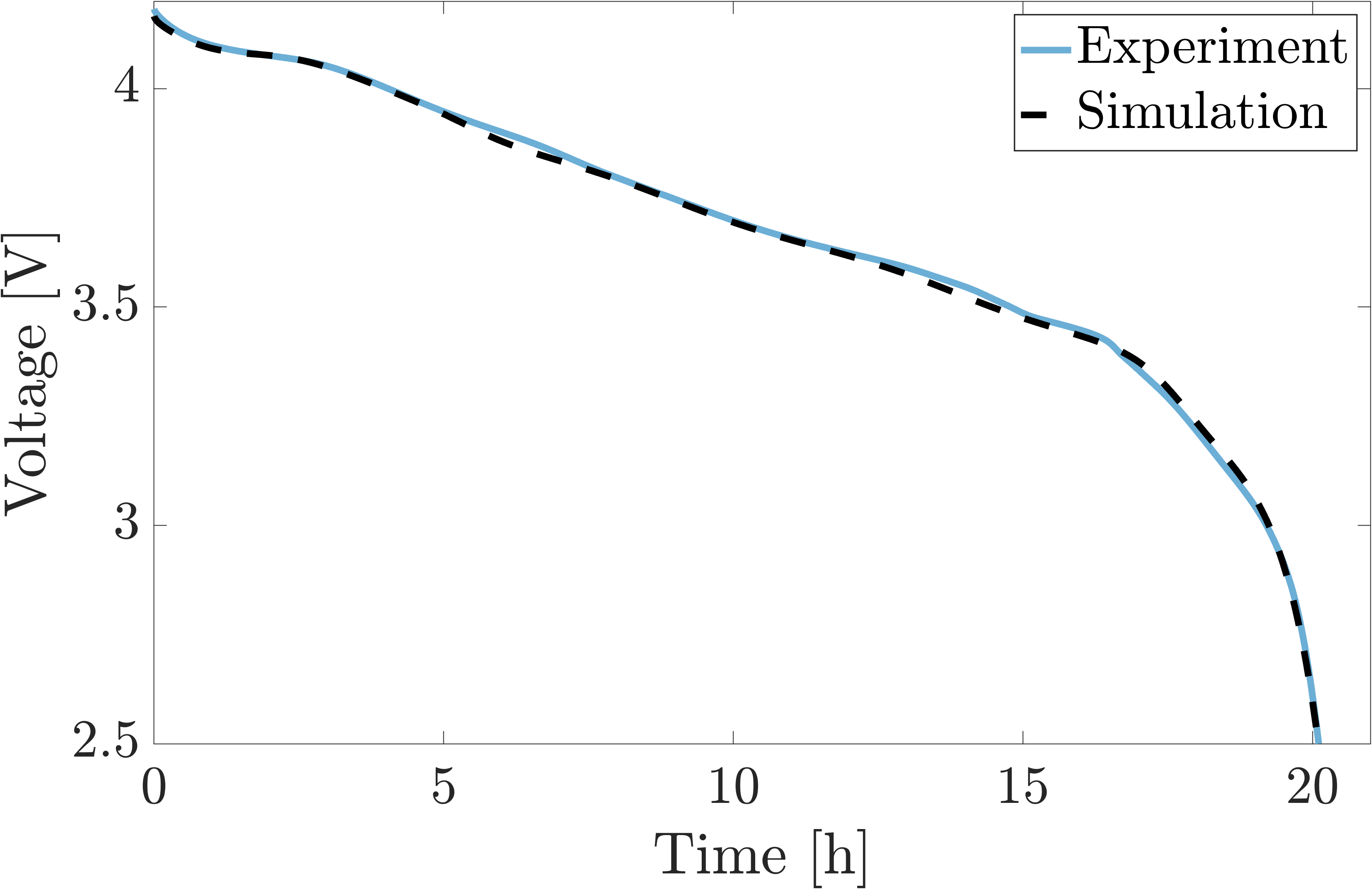}
\caption{C/20 capacity test voltage identification
results.\label{fig:V_0_05C}}
\end{figure}

\begin{figure}
\centering
\includegraphics[width=0.65\textwidth,height=\textheight]{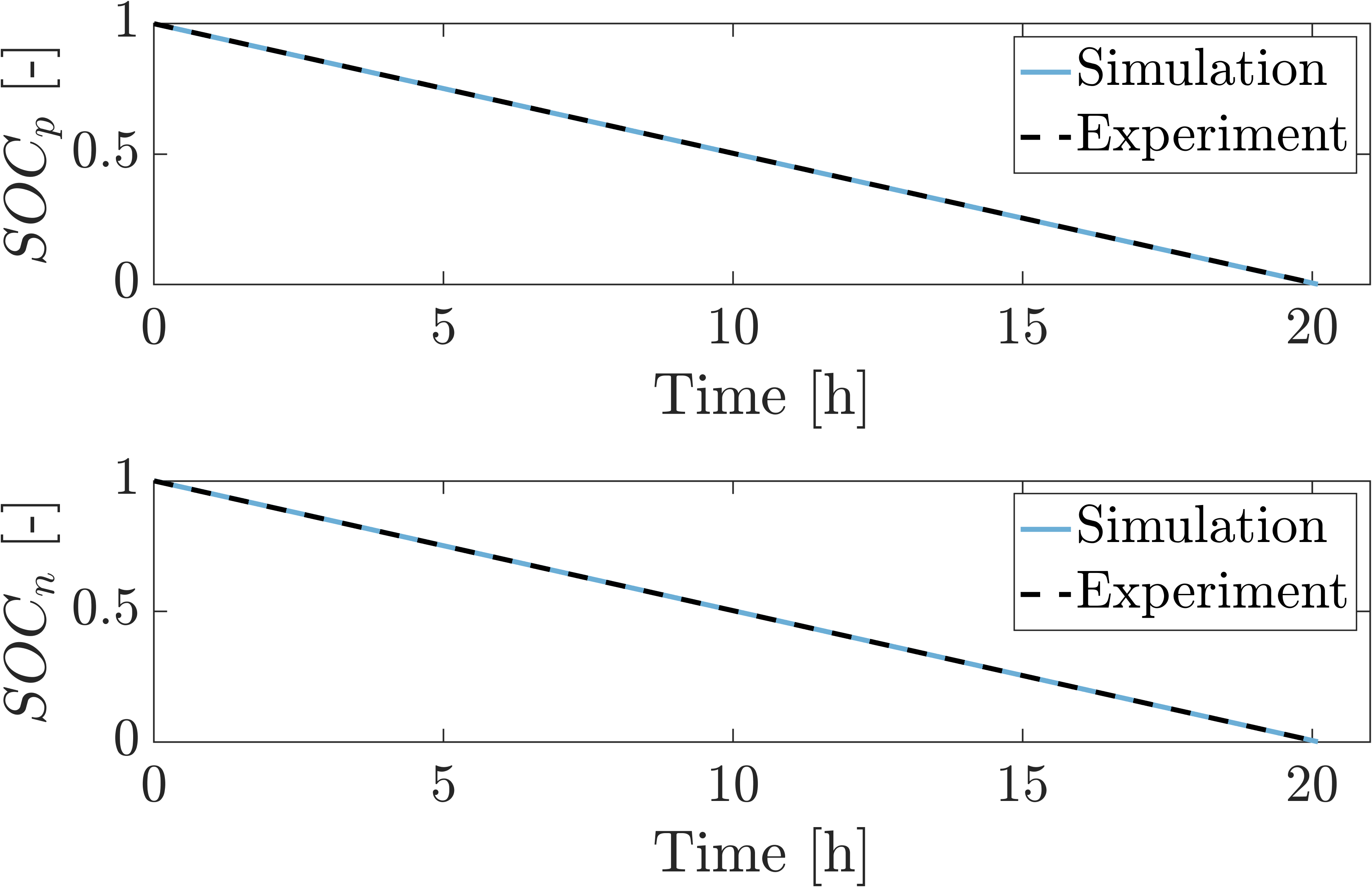}
\caption{C/20 capacity test positive and negative electrode SOC
identification results.\label{fig:SOC_0_05C}}
\end{figure}

The code also plots the simulation results generated from the identified
parameters and the experimental data, as shown in \autoref{fig:V_0_05C}
and \autoref{fig:SOC_0_05C}.

Run
\texttt{Examples/Parameter\_Identification\_Results/DFN\_pso\_0\_05C\_identification.m}
to view the C/20 identification results shown here.

\hypertarget{hppc-identification}{%
\subsection{HPPC Identification}\label{hppc-identification}}

The \texttt{DFN\_pso\_HPPC.m} file's \texttt{User\ Input} section is
similar to the one described in \texttt{DFN\_pso\_0\_05C.m}. First, load
your \texttt{param} structure, which contains the nominal DFN parameters
and any previously identified parameter values. In this example, we load
the \texttt{identified\_parameters\_0\_05C.mat} file, which contains
stoichiometric parameter identification results:

\begin{Shaded}
\begin{Highlighting}[]
\CommentTok{\%\% User Input}
\CommentTok{\% Load nominal parameters and identified stoichiometric parameters}
\CommentTok{\% from C/20 capacity test data}
\VariableTok{load}\NormalTok{(}\SpecialStringTok{\textquotesingle{}identified\_parameters\_0\_05C.mat\textquotesingle{}}\OperatorTok{,}\SpecialStringTok{\textquotesingle{}param\textquotesingle{}}\NormalTok{)}
\end{Highlighting}
\end{Shaded}

When defining the names of the HPPC parameters to identify, users can
manually type the parameters (Option 1) or load the parameter
identifiability results generated from \texttt{DFN\_LSA\_Corr\_HPPC.m}
(Option 2).

In Option 1, all the unknown transport and kinetic parameters are
identified, consisting of the reaction rate constants in the electrodes
\(k_p\) (\texttt{kp}) and \(k_n\) (\texttt{kn}), electrolyte
diffusitivity \(D_e\) (\texttt{De}), electrolyte conductivity \(\kappa\)
(\texttt{Kappa}), transference number \(t_+\) (\texttt{t1\_constant}),
initial electrolyte concentration (\texttt{c0}), and solid phase
diffusitivities \(D_{s,p}\) (\texttt{Dsp}) and \(D_{s,n}\)
(\texttt{Dsn}):

\begin{Shaded}
\begin{Highlighting}[]
\CommentTok{\%{-}{-}{-}{-}{-}{-}{-}{-}{-}{-}{-}{-}{-}{-}{-}{-}{-}{-}{-}{-}{-}{-}{-}{-}{-}{-}{-}{-}{-}{-}{-}{-}{-}{-}{-}{-}{-}{-}{-}{-}{-}{-}{-}{-}{-}{-}{-}{-}{-}{-}{-}{-}{-}{-}{-}{-}{-}{-}{-}{-}{-}{-}{-}{-}{-}{-}{-}{-}{-}{-}{-}{-}{-}{-}}
\CommentTok{\% Option 1: Enter names of parameters to identify (make sure names match the}
\CommentTok{\% parameter names in "param" structure containing nominal parameters)}
\CommentTok{\%{-}{-}{-}{-}{-}{-}{-}{-}{-}{-}{-}{-}{-}{-}{-}{-}{-}{-}{-}{-}{-}{-}{-}{-}{-}{-}{-}{-}{-}{-}{-}{-}{-}{-}{-}{-}{-}{-}{-}{-}{-}{-}{-}{-}{-}{-}{-}{-}{-}{-}{-}{-}{-}{-}{-}{-}{-}{-}{-}{-}{-}{-}{-}{-}{-}{-}{-}{-}{-}{-}{-}{-}{-}{-}}
\VariableTok{param\_HPPC} \OperatorTok{=}\NormalTok{ \{}\SpecialStringTok{\textquotesingle{}Dsp\textquotesingle{}} \SpecialStringTok{\textquotesingle{}Dsn\textquotesingle{}} \SpecialStringTok{\textquotesingle{}t1\_constant\textquotesingle{}} \SpecialStringTok{\textquotesingle{}kp\textquotesingle{}} \SpecialStringTok{\textquotesingle{}kn\textquotesingle{}} \SpecialStringTok{\textquotesingle{}c0\textquotesingle{}} \SpecialStringTok{\textquotesingle{}De\textquotesingle{}} \SpecialStringTok{\textquotesingle{}Kappa\textquotesingle{}}\NormalTok{\}}\OperatorTok{;}
\end{Highlighting}
\end{Shaded}

Option 2 uses identifiability analysis results from
\texttt{DFN\_LSA\_Corr\_HPPC.m}, which produces two sets of parameters:
\texttt{LSA\_identifiable} and \texttt{corr\_identifiable}. The former
includes parameters that have sensitivities higher than the user-defined
threshold (\texttt{beta\_LSA}). The latter consists of parameters with
high sensivity and correlation coefficients lower than the specified
correlation threshold (\texttt{beta\_corr}).

In this example, identification of the \texttt{LSA\_identifiable}
parameter set is investigated:

\begin{Shaded}
\begin{Highlighting}[]
\CommentTok{\%{-}{-}{-}{-}{-}{-}{-}{-}{-}{-}{-}{-}{-}{-}{-}{-}{-}{-}{-}{-}{-}{-}{-}{-}{-}{-}{-}{-}{-}{-}{-}{-}{-}{-}{-}{-}{-}{-}{-}{-}{-}{-}{-}{-}{-}{-}{-}{-}{-}{-}{-}{-}{-}{-}{-}{-}{-}{-}{-}{-}{-}{-}{-}{-}{-}{-}{-}{-}{-}{-}{-}{-}{-}{-}}
\CommentTok{\% Option 2: Load identifiable parameters from identifiability analysis}
\CommentTok{\% conducted in "Examples/Local\_Sensitivity\_Analysis/DFN\_LSA\_Corr\_HPPC.m"}
\CommentTok{\%{-}{-}{-}{-}{-}{-}{-}{-}{-}{-}{-}{-}{-}{-}{-}{-}{-}{-}{-}{-}{-}{-}{-}{-}{-}{-}{-}{-}{-}{-}{-}{-}{-}{-}{-}{-}{-}{-}{-}{-}{-}{-}{-}{-}{-}{-}{-}{-}{-}{-}{-}{-}{-}{-}{-}{-}{-}{-}{-}{-}{-}{-}{-}{-}{-}{-}{-}{-}{-}{-}{-}{-}{-}{-}}
\CommentTok{\% \textquotesingle{}LSA\_identifiable\textquotesingle{} {-}\textgreater{} parameters with sensitivity higher than beta\_LSA}
\CommentTok{\% \textquotesingle{}corr\_identifiable\textquotesingle{} {-}\textgreater{} parameters determined through corr. analysis with}
\CommentTok{\%                        a corr. threshold of beta\_corr}
\CommentTok{\%{-}{-}{-}{-}{-}{-}{-}{-}{-}{-}{-}{-}{-}{-}{-}{-}{-}{-}{-}{-}{-}{-}{-}{-}{-}{-}{-}{-}{-}{-}{-}{-}{-}{-}{-}{-}{-}{-}{-}{-}{-}{-}{-}{-}{-}{-}{-}{-}{-}{-}{-}{-}{-}{-}{-}{-}{-}{-}{-}{-}{-}{-}{-}{-}{-}{-}{-}{-}{-}{-}{-}{-}{-}{-}}
\VariableTok{load}\NormalTok{(}\SpecialStringTok{\textquotesingle{}DFN\_identification\_results/HPPC\_identifiable\_params.mat\textquotesingle{}}\OperatorTok{,...}
    \SpecialStringTok{\textquotesingle{}LSA\_identifiable\textquotesingle{}}\OperatorTok{,}\SpecialStringTok{\textquotesingle{}corr\_identifiable\textquotesingle{}}\NormalTok{)}
\VariableTok{param\_HPPC} \OperatorTok{=} \VariableTok{LSA\_identifiable}\OperatorTok{;}
\end{Highlighting}
\end{Shaded}

Enter the MAT file name to save an updated param structure containing
the PSO results:

\begin{Shaded}
\begin{Highlighting}[]
\CommentTok{\% Enter mat file name where your PSO results will be stored}
\VariableTok{file\_name} \OperatorTok{=} \SpecialStringTok{\textquotesingle{}identified\_parameters\_HPPC\_noCorr\textquotesingle{}}\OperatorTok{;}
\end{Highlighting}
\end{Shaded}

Define the upper and lower bounds for each parameter in
\texttt{param\_HPPC}:

\begin{Shaded}
\begin{Highlighting}[]
\CommentTok{\% Enter lower and upper bounds of parameters to identify }
\CommentTok{\% Dsp}
\VariableTok{pct} \OperatorTok{=} \FloatTok{0.2}\OperatorTok{;} \CommentTok{\% perturbation coeff}
\VariableTok{lower\_bounds}\NormalTok{.}\VariableTok{Dsp} \OperatorTok{=} \FloatTok{10}\OperatorTok{\^{}}\NormalTok{(}\VariableTok{log10}\NormalTok{(}\VariableTok{param}\NormalTok{.}\VariableTok{Dsp}\NormalTok{)}\OperatorTok{*}\NormalTok{(}\FloatTok{1}\OperatorTok{+}\VariableTok{pct}\NormalTok{))}\OperatorTok{;}
\VariableTok{upper\_bounds}\NormalTok{.}\VariableTok{Dsp} \OperatorTok{=} \FloatTok{10}\OperatorTok{\^{}}\NormalTok{(}\VariableTok{log10}\NormalTok{(}\VariableTok{param}\NormalTok{.}\VariableTok{Dsp}\NormalTok{)}\OperatorTok{*}\NormalTok{(}\FloatTok{1}\OperatorTok{{-}}\VariableTok{pct}\NormalTok{))}\OperatorTok{;}
\OperatorTok{...}
\end{Highlighting}
\end{Shaded}

Load the time, current, and voltage vectors generated from the HPPC
data:

\begin{Shaded}
\begin{Highlighting}[]
\CommentTok{\% Load Experimental Data }
\CommentTok{\% HPPC test conducted on LG INR21700 M50T cells}
\VariableTok{load}\NormalTok{(}\SpecialStringTok{\textquotesingle{}data\_INR21700\_M50T/HPPC\_data\_W8\_Diag1.mat\textquotesingle{}}\NormalTok{)    }
\end{Highlighting}
\end{Shaded}

Once all user inputs has been defined, run the code to start the PSO.
Once the PSO is completed, the identified parameter and objective
function values are printed to the Command Window:

\begin{verbatim}
Displaying identified values...
------------------------
Dsp:
Identified value: 5.5423e-15
5.278e-18(lower) | 4e-15(initial) | 3.0314e-12(upper)
------------------------
kn:
Identified value: 8.6471e-09
2.987e-15(lower) | 6.7159e-12(initial) | 1.51e-08(upper)
------------------------
c0:
Identified value: 1166.3688
500(lower) | 1000(initial) | 1500(upper)
------------------------
Dsn:
Identified value: 2.1618e-14
6.6407e-17(lower) | 3.3e-14(initial) | 1.6399e-11(upper)

Displaying objective function values...
------------------------
J_V =0.0038222 [-]
J_V =13.4785 [mV]
J_SOCp =0.13299 [%]
J_SOCn =0.17276 [%]
J_tot =0.0068797 [-]
\end{verbatim}

\begin{figure}
\centering
\includegraphics[width=0.65\textwidth,height=\textheight]{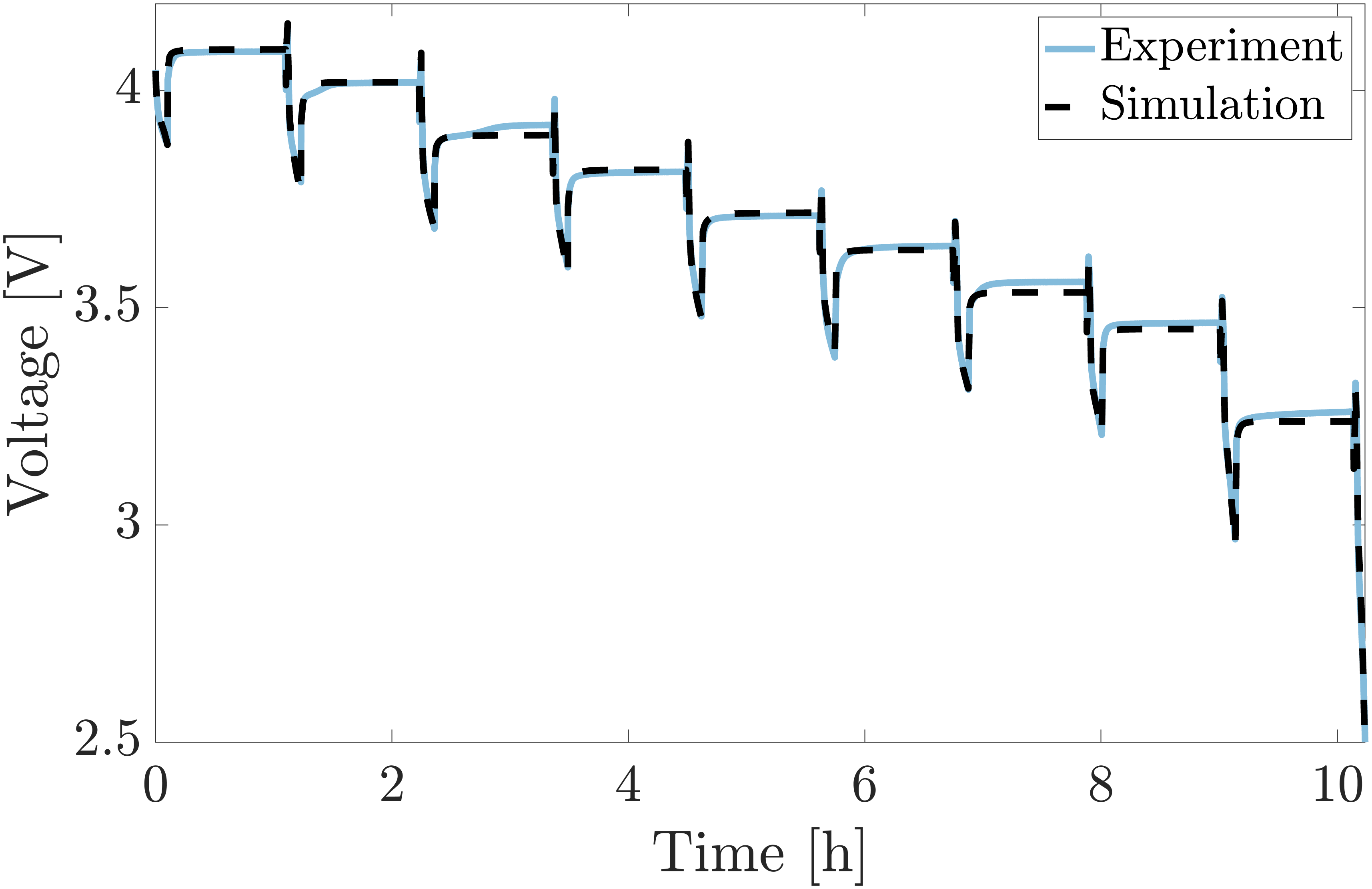}
\caption{HPPC voltage identification results.\label{fig:V_HPPC}}
\end{figure}

\begin{figure}
\centering
\includegraphics[width=0.65\textwidth,height=\textheight]{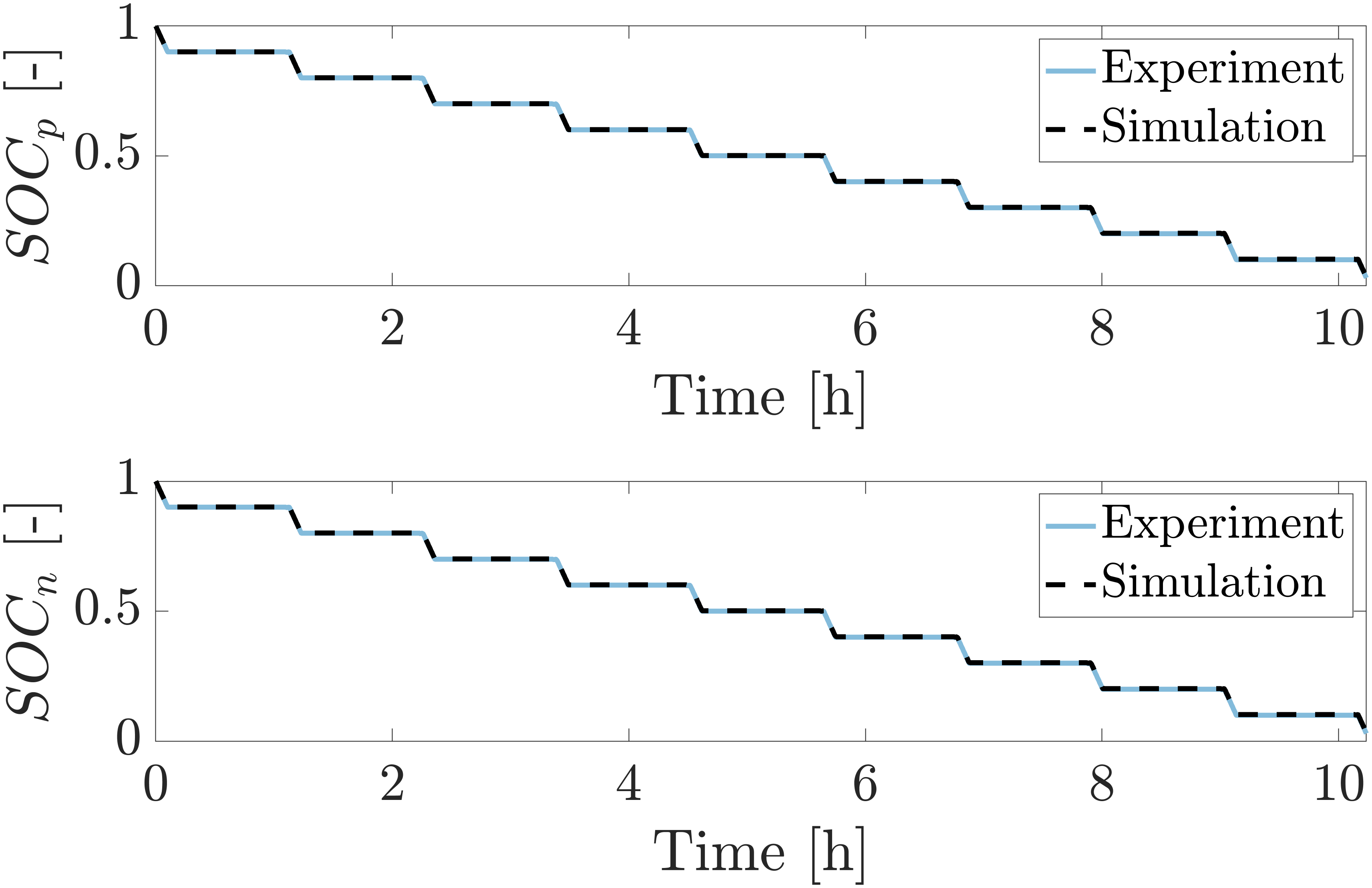}
\caption{HPPC positive and negative electrode SOC identification
results.\label{fig:SOC_HPPC}}
\end{figure}

Similar to \texttt{DFN\_pso\_0\_05C.m}, the simulation results generated
from the identified parameters are plotted against the experimental
data, as shown in \autoref{fig:V_HPPC} and \autoref{fig:SOC_HPPC}.

Run
\texttt{Examples/Parameter\_Identification\_Results/DFN\_pso\_HPPC\_identification.m}
to view the HPPC identification results shown here.

\hypertarget{udds-driving-cycle-validation}{%
\subsection{UDDS Driving Cycle
Validation}\label{udds-driving-cycle-validation}}

In the code
\texttt{Examples/Parameter\_Identification\_Results/DFN\_pso\_UDDS\_validation.m},
the identified parameters from the C/20 capacity test and HPPC data are
validated using the UDDS driving cycle. The model is simulated under the
UDDS profile and compared against the experimental UDDS data.

In the \texttt{User\ Input} section, load the parameter values
identified from C/20 and HPPC data:

\begin{Shaded}
\begin{Highlighting}[]
\CommentTok{\%\% User Input  }
\CommentTok{\% Load identification results }
\VariableTok{load}\NormalTok{(}\SpecialStringTok{\textquotesingle{}identified\_parameters\_HPPC\_noCorr.mat\textquotesingle{}}\OperatorTok{,}\SpecialStringTok{\textquotesingle{}param\textquotesingle{}}\NormalTok{)}
\end{Highlighting}
\end{Shaded}

and load the experimental UDDS data:

\begin{Shaded}
\begin{Highlighting}[]
\CommentTok{\% Load Experimental Data }
\CommentTok{\% HPPC test conducted on LG INR21700 M50T cells}
\VariableTok{load}\NormalTok{(}\SpecialStringTok{\textquotesingle{}data\_INR21700\_M50T/UDDS\_W8\_cyc1.mat\textquotesingle{}}\NormalTok{)}
\end{Highlighting}
\end{Shaded}

\begin{figure}
\centering
\includegraphics[width=0.65\textwidth,height=\textheight]{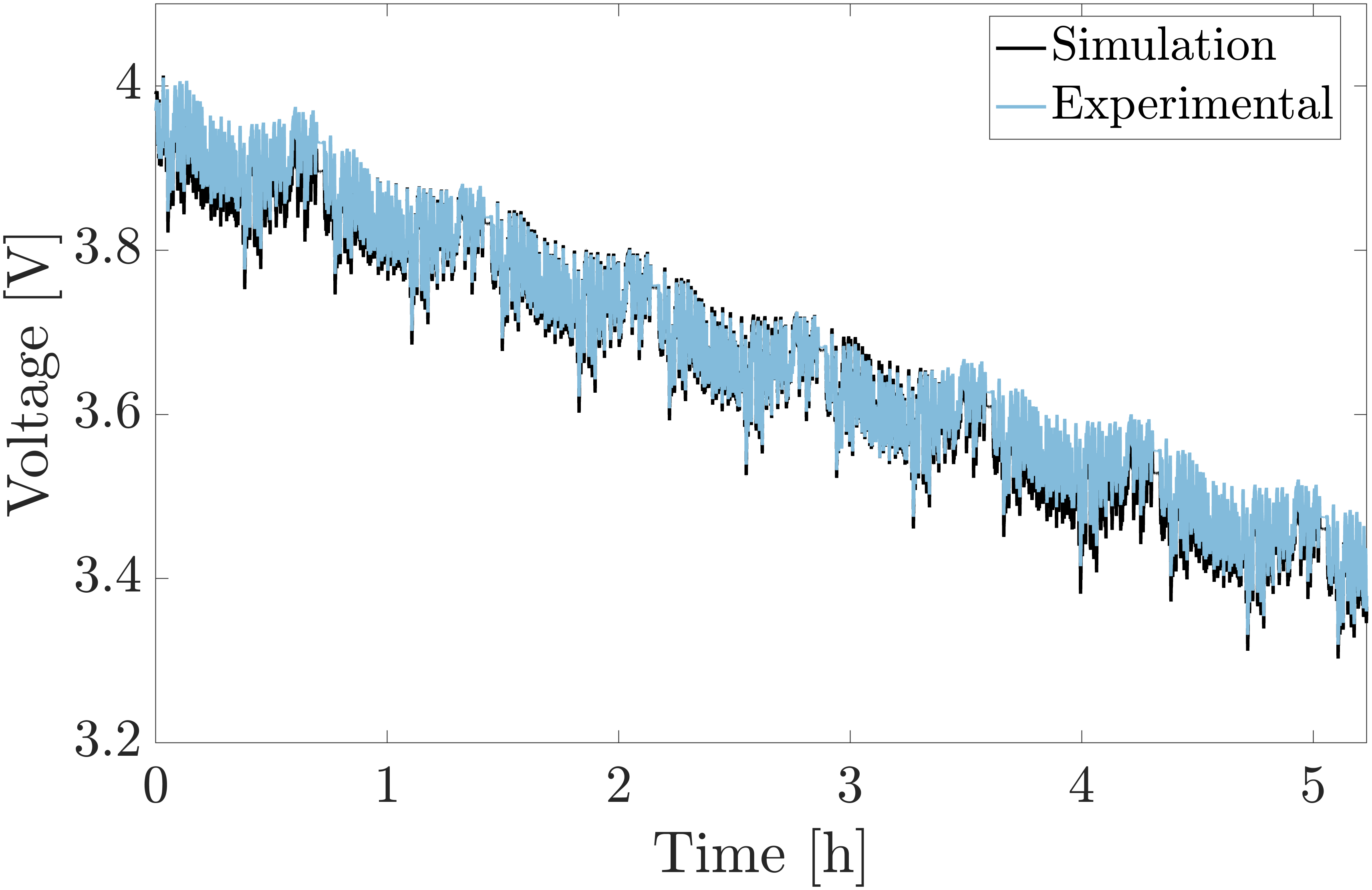}
\caption{UDDS voltage identification results.\label{fig:V_UDDS}}
\end{figure}

\begin{figure}
\centering
\includegraphics[width=0.65\textwidth,height=\textheight]{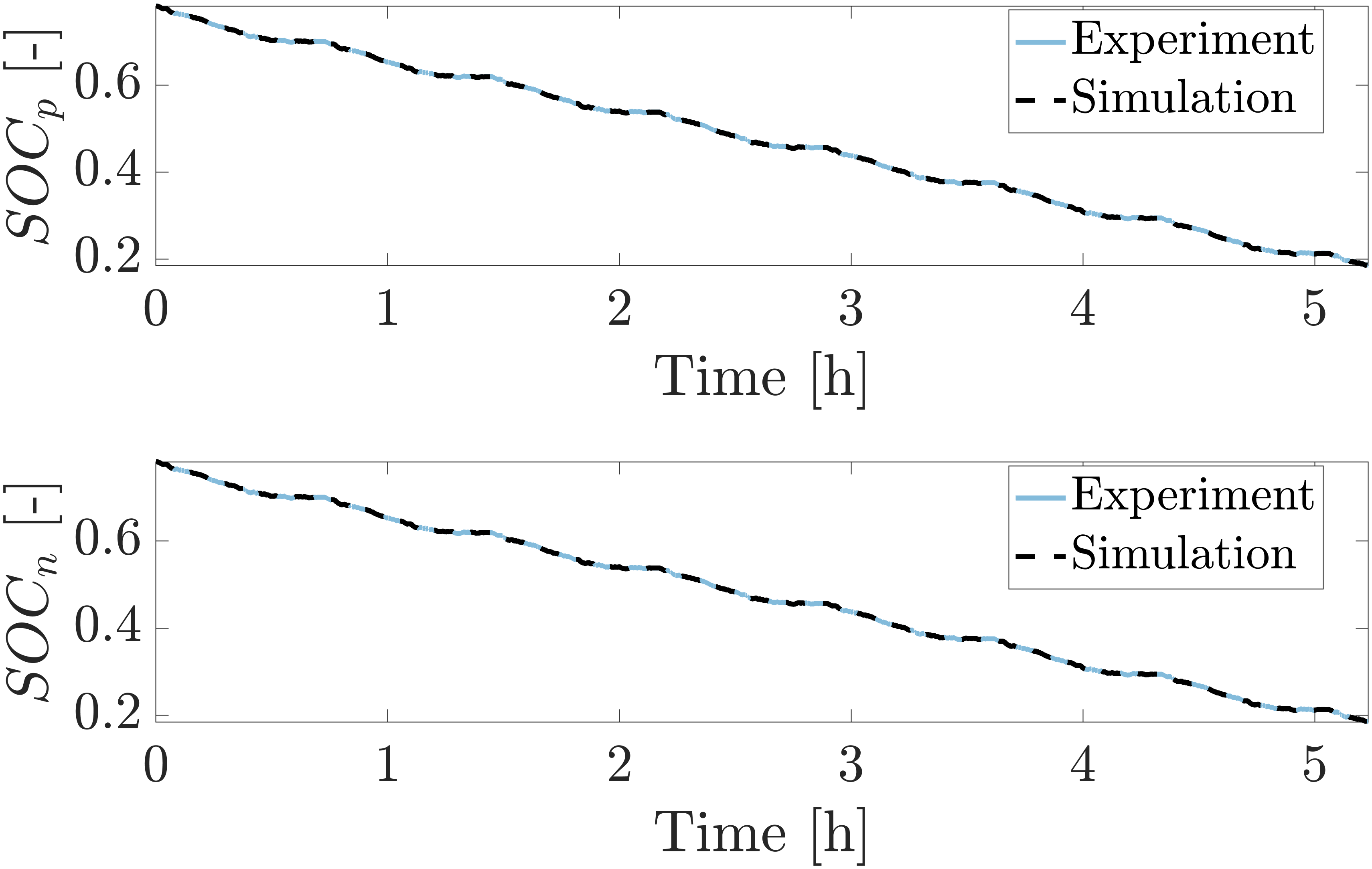}
\caption{UDDS positive and negative electrode SOC identification
results.\label{fig:SOC_UDDS}}
\end{figure}

The objective function is printed to the Command Window:

\begin{verbatim}
Displaying objective function values...
------------------------
J_V =0.0039168 [-]
J_V =14.3911 [mV]
J_SOCp =0.032573 [%]
J_SOCn =0.015161 [%]
J_tot =0.0043941 [-]
\end{verbatim}

The simulation results and experimental data are plotted as shown in
\autoref{fig:V_UDDS} and \autoref{fig:SOC_UDDS}.

Visit COBRAPRO's
\href{https://github.com/COBRAPROsimulator/COBRAPRO}{Github} to view all
example codes:

\begin{itemize}
\tightlist
\item
  \texttt{Examples/Parameter\_Identification\_Routines}: Parameter
  identification examples

  \begin{itemize}
  \tightlist
  \item
    \texttt{DFN\_pso\_0\_05C.m}: Parameter identification using C/20
    capacity test data
  \item
    \texttt{DFN\_pso\_HPPC.m}: Parameter identification using HPPC data
  \end{itemize}
\item
  \texttt{Examples/Parameter\_Identification\_Results}: Load parameter
  identification results

  \begin{itemize}
  \tightlist
  \item
    \texttt{DFN\_pso\_0\_05C\_identification.m}: C/20 capacity test
    identification results
  \item
    \texttt{DFN\_pso\_HPPC\_identification.m}: HPPC identification
    results
  \item
    \texttt{DFN\_pso\_UDDS\_validation.m}: UDDS validation results
  \end{itemize}
\item
  \texttt{Examples/Cycling}: Simulating battery cycling examples

  \begin{itemize}
  \tightlist
  \item
    \texttt{cycle\_CC.m}: CC cycling and model output visualization
  \item
    \texttt{cycle\_HPPC.m}: HPPC simulation and model output
    visualization
  \item
    \texttt{cycle\_UDDS.m}: UDDS simulation and model output
    visualization
  \end{itemize}
\item
  \texttt{Examples/Local\_Sensitivity\_Analysis}: LSA and correlation
  analysis examples

  \begin{itemize}
  \tightlist
  \item
    \texttt{DFN\_LSA\_Corr\_CC.m}: LSA and correlation analysis on CC
    profile
  \item
    \texttt{DFN\_LSA\_Corr\_HPPC.m}: LSA and correlation analysis on
    HPPC profile
  \end{itemize}
\end{itemize}

\hypertarget{acknowledgements}{%
\section{Acknowledgements}\label{acknowledgements}}

The authors thank the Bits and Watts Initiative within the Precourt
Institute for Energy at Stanford University for its partial financial
support. We thank Dr.~Le Xu for all the insightful discussions that
greatly contributed to the enhancement of COBRAPRO. We extend our thanks
to Alexis Geslin, Joseph Lucero, and Maitri Uppaluri for testing
COBRAPRO and providing valuable feedback.

\hypertarget{references}{%
\section*{References}\label{references}}
\addcontentsline{toc}{section}{References}

\hypertarget{refs}{}
\begin{CSLReferences}{1}{0}
\leavevmode\vadjust pre{\hypertarget{ref-arunachalam_full_2019}{}}%
Arunachalam, H., \& Onori, S. (2019). Full {Homogenized Macroscale
Model} and {Pseudo-2-Dimensional Model} for {Lithium-Ion Battery
Dynamics}: {Comparative Analysis}, {Experimental Verification} and
{Sensitivity Analysis}. \emph{Journal of The Electrochemical Society},
\emph{166}(8), A1380--A1392. \url{https://doi.org/10.1149/2.0051908jes}

\leavevmode\vadjust pre{\hypertarget{ref-berliner_methods_2021}{}}%
Berliner, M. D., Cogswell, D. A., Bazant, M. Z., \& Braatz, R. D.
(2021). Methods{\textemdash}{PETLION}: {Open-Source Software} for
{Millisecond-Scale Porous Electrode Theory-Based Lithium-Ion Battery
Simulations}. \emph{Journal of The Electrochemical Society},
\emph{168}(9), 090504. \url{https://doi.org/10.1149/1945-7111/ac201c}

\leavevmode\vadjust pre{\hypertarget{ref-chen_development_2020}{}}%
Chen, C.-H., Brosa Planella, F., O'Regan, K., Gastol, D., Widanage, W.
D., \& Kendrick, E. (2020). Development of {Experimental Techniques} for
{Parameterization} of {Multi-scale Lithium-ion Battery Models}.
\emph{Journal of The Electrochemical Society}, \emph{167}(8), 080534.
\url{https://doi.org/10.1149/1945-7111/ab9050}

\leavevmode\vadjust pre{\hypertarget{ref-comsol}{}}%
COMSOL AB, Stockholm, Sweden. (2023). \emph{COMSOL multiphysics\&copy;
v. 6.2}. \href{https://www.comsol.com}{www.comsol.com}

\leavevmode\vadjust pre{\hypertarget{ref-couto_lithiumion_2023}{}}%
Couto, Luis. D., Charkhgard, M., Karaman, B., Job, N., \& Kinnaert, M.
(2023). Lithium-ion battery design optimization based on a dimensionless
reduced-order electrochemical model. \emph{Energy}, \emph{263}, 125966.
\url{https://doi.org/10.1016/j.energy.2022.125966}

\leavevmode\vadjust pre{\hypertarget{ref-dai_graded_2016}{}}%
Dai, Y., \& Srinivasan, V. (2016). On {Graded Electrode Porosity} as a
{Design Tool} for {Improving} the {Energy Density} of {Batteries}.
\emph{Journal of The Electrochemical Society}, \emph{163}(3),
A406--A416. \url{https://doi.org/10.1149/2.0301603jes}

\leavevmode\vadjust pre{\hypertarget{ref-doyle_modeling_1993}{}}%
Doyle, M., Fuller, T. F., \& Newman, J. (1993). Modeling of
{Galvanostatic Charge} and {Discharge} of the
{Lithium}/{Polymer}/{Insertion Cell}. \emph{Journal of The
Electrochemical Society}, \emph{140}(6), 1526--1533.
\url{https://doi.org/10.1149/1.2221597}

\leavevmode\vadjust pre{\hypertarget{ref-ecker_parameterization_2015a}{}}%
Ecker, M., Tran, T. K. D., Dechent, P., Käbitz, S., Warnecke, A., \&
Sauer, D. U. (2015). Parameterization of a {Physico-Chemical Model} of a
{Lithium-Ion Battery}: {I}. {Determination} of {Parameters}.
\emph{Journal of The Electrochemical Society}, \emph{162}(9),
A1836--A1848. \url{https://doi.org/10.1149/2.0551509jes}

\leavevmode\vadjust pre{\hypertarget{ref-kolluri_realtime_2020}{}}%
Kolluri, S., Aduru, S. V., Pathak, M., Braatz, R. D., \& Subramanian, V.
R. (2020). Real-time {Nonlinear Model Predictive Control} ({NMPC})
{Strategies} using {Physics-Based Models} for {Advanced Lithium-ion
Battery Management System} ({BMS}). \emph{Journal of The Electrochemical
Society}, \emph{167}(6), 063505.
\url{https://doi.org/10.1149/1945-7111/ab7bd7}

\leavevmode\vadjust pre{\hypertarget{ref-lawder_extending_2015}{}}%
Lawder, M. T., Ramadesigan, V., Suthar, B., \& Subramanian, V. R.
(2015). Extending explicit and linearly implicit {ODE} solvers for
index-1 {DAEs}. \emph{Computers \& Chemical Engineering}, \emph{82},
283--292. \url{https://doi.org/10.1016/j.compchemeng.2015.07.002}

\leavevmode\vadjust pre{\hypertarget{ref-lee_robust_2021}{}}%
Lee, S. B., \& Onori, S. (2021). A {Robust} and {Sleek Electrochemical
Battery Model Implementation}: {A MATLAB}{\textregistered} {Framework}.
\emph{Journal of The Electrochemical Society}, \emph{168}(9), 090527.
\url{https://doi.org/10.1149/1945-7111/ac22c8}

\leavevmode\vadjust pre{\hypertarget{ref-methekar_perturbation_2011}{}}%
Methekar, R. N., Ramadesigan, V., Pirkle, J. C., \& Subramanian, V. R.
(2011). A perturbation approach for consistent initialization of index-1
explicit differential{\textendash}algebraic equations arising from
battery model simulations. \emph{Computers \& Chemical Engineering},
\emph{35}(11), 2227--2234.
\url{https://doi.org/10.1016/j.compchemeng.2011.01.003}

\leavevmode\vadjust pre{\hypertarget{ref-pozzato_data_2022}{}}%
Pozzato, G., Allam, A., \& Onori, S. (2022). Lithium-ion battery aging
dataset based on electric vehicle real-driving profiles. \emph{Data in
Brief}, \emph{41}, 107995.
\url{https://doi.org/10.1016/j.dib.2022.107995}

\leavevmode\vadjust pre{\hypertarget{ref-pozzato_general_2023}{}}%
Pozzato, G., \& Onori, S. (2023). A {General Matlab} and {COMSOL
Co-simulation Framework} for {Model Parameter Optimization}:
{Lithium-Ion Battery} and {Gasoline Particulate Filter Case Studies}.
\emph{Automotive {Technical Papers}}, 2023-01-5047.
\url{https://doi.org/10.4271/2023-01-5047}

\leavevmode\vadjust pre{\hypertarget{ref-schmalstieg_full_2018a}{}}%
Schmalstieg, J., Rahe, C., Ecker, M., \& Sauer, D. U. (2018). Full {Cell
Parameterization} of a {High-Power Lithium-Ion Battery} for a
{Physico-Chemical Model}: {Part I}. {Physical} and {Electrochemical
Parameters}. \emph{Journal of The Electrochemical Society},
\emph{165}(16), A3799--A3810. \url{https://doi.org/10.1149/2.0321816jes}

\leavevmode\vadjust pre{\hypertarget{ref-fastDFN}{}}%
Scott Moura. (2016). \emph{fastDFN}.
\url{https://github.com/scott-moura/fastDFN}

\leavevmode\vadjust pre{\hypertarget{ref-smith_multiphase_2017}{}}%
Smith, R. B., \& Bazant, M. Z. (2017). Multiphase {Porous Electrode
Theory}. \emph{Journal of The Electrochemical Society}, \emph{164}(11),
E3291--E3310. \url{https://doi.org/10.1149/2.0171711jes}

\leavevmode\vadjust pre{\hypertarget{ref-sulzer_python_2021}{}}%
Sulzer, V., Marquis, S. G., Timms, R., Robinson, M., \& Chapman, S. J.
(2021). Python {Battery Mathematical Modelling} ({PyBaMM}).
\emph{Journal of Open Research Software}, \emph{9}(1), 14.
\url{https://doi.org/10.5334/jors.309}

\leavevmode\vadjust pre{\hypertarget{ref-torchio_lionsimba_2016}{}}%
Torchio, M., Magni, L., Gopaluni, R. B., Braatz, R. D., \& Raimondo, D.
M. (2016). {LIONSIMBA}: {A Matlab Framework Based} on a {Finite Volume
Model Suitable} for {Li-Ion Battery Design}, {Simulation}, and
{Control}. \emph{Journal of The Electrochemical Society}, \emph{163}(7),
A1192--A1205. \url{https://doi.org/10.1149/2.0291607jes}

\leavevmode\vadjust pre{\hypertarget{ref-xu_comparative_2023}{}}%
Xu, L., Cooper, J., Allam, A., \& Onori, S. (2023). Comparative
{Analysis} of {Numerical Methods} for {Lithium-Ion Battery
Electrochemical Modeling}. \emph{Journal of The Electrochemical
Society}, \emph{170}(12), 120525.
\url{https://doi.org/10.1149/1945-7111/ad1293}

\end{CSLReferences}

\end{document}